# FIRST RESULTS FROM NANOINDENTATION OF VAPOR DIFFUSED Nb$_3$Sn FILMS ON Nb

U. Pudasaini[†1], G.V. Eremeev[2], S. Cheban[2]
[1]Thomas Jefferson National Accelerator Facility, Newport News, VA 23606, USA
[2]Fermi National Accelerator Laboratory, Batavia, IL 60510, USA

## Abstract

The mechanical vulnerability of the Nb$_3$Sn-coated cavities is identified as one of the significant technical hurdles toward deploying them in practical accelerator applications in the not-so-distant future. It is crucial to characterize the material's mechanical properties in ways to address such vulnerability. Nanoindentation is a widely used technique for measuring the mechanical properties of thin films that involves indenting the film with a small diamond tip and measuring the force-displacement response to calculate the film's elastic modulus, hardness, and other mechanical properties. The nanoindentation analysis was performed on multiple vapor-diffused Nb3Sn samples coated at Jefferson Lab and Fermilab coating facilities for the first time. This contribution will discuss the first results obtained from the nanoindentation of Nb$_3$Sn-coated Nb samples prepared via the Sn vapor diffusion technique.

## INTRODUCTION

Nb$_3$Sn, with a superconducting transition temperature of ~18.2 K and a superheating field of ~400 mT, is a leading alternative material to replace niobium in SRF accelerator cavities [1]. Accordingly, it promises a higher accelerating gradient, quality factor, and operation temperature than traditional bulk Nb. Operating Nb$_3$Sn SRF cavities at 4.3 K can deliver similar performance to Nb cavities at 2 K, resulting in enormous cost savings for SRF accelerators. That means these cavities can be operated with atmospheric liquid helium or cryocoolers, simplifying and reducing the cost of cryogenic facilities. The successful deployment of Nb$_3$Sn technology will be transformational, significantly benefiting numerous SRF accelerators and enabling new classes of SRF accelerator applications.

Since Nb$_3$Sn is a very brittle material with a significantly lower thermal conductivity than Nb, it should be grown as a thin film for application. Several alternate coating techniques are being pursued at multiple labs to grow and optimize Nb$_3$Sn thin film on metallic structures. Still, the Sn vapor diffusion process is yet the more mature technique for conformality and the only one thus far that has produced rf results for Nb$_3$Sn-coated Nb cavities. The state-of-the-art single-cell Nb$_3$Sn cavity frequently attains accelerating gradients of $\geq$ 15MV/m with a quality factor $\geq 10^{10}$ [2-5]. Several Nb$_3$Sn-coated multi-cell cavities have reached ~15 MV with a quality factor of ~$10^{10}$ [4, 6]. A significant improvement has been made in the performance of Nb$_3$Sn-coated cavities over the last decade; these cavities are already suitable for some accelerator applications. Several projects in different laboratories are considering Nb$_3$Sn-coated cavities for small accelerator applications. The construction of a quarter module using two CEBAF-style C75 cavities is in the final stage at Jefferson Lab. The quarter cryomodule will be installed in the upgraded injector test facility (UITF) to accelerate an electron beam up to 10 MeV [7]. If successful, the facility can use a cryomodule with Nb$_3$Sn-coated cavities to run low-energy nuclear physics experiments at 4 K. Nb$_3$Sn cavities have the potential to enable further and significantly simplify widespread use of SRF technology in light-source storage rings, FELs, and other compact accelerators. There have been successful tests of Nb$_3$Sn cavities operating in conduction-cooled setups as demonstrations suitable for industrial accelerator applications at Fermilab (650 MHz single cell cavity), JLab (1.5 GHz and 952 MHz single cell), and Cornell (2.6 GHz) [8-10]. Detailed plans have been published for a medium-energy, high average-power superconducting e-beam accelerator for environmental applications at Fermilab [11] and a CW, low-energy, high-power superconducting linac for environmental applications by researchers at JLab [12].

Because of the material's brittleness, the mechanical vulnerability is identified as a significant technical challenge in deploying the Nb$_3$Sn-coated cavities in practical accelerators. The performance degradation of a Nb$_3$Sn-coated cavity resulting from the tuning of ~300 KHz at room temperatures has been demonstrated [13]. To address this challenge, it is essential to understand the mechanical properties and behavior of vapor-diffused Nb$_3$Sn thin film. So far, per the authors' knowledge, no such studies have been reported before; we used the nanoindentation technique to obtain fundamental mechanical properties such as elastic modulus, hardness, and yield stress. In this contribution, the first results from nanoindentation of vapor-diffused Nb$_3$Sn coatings on differently prepared Nb substrates coated in Fermilab and Jefferson Lab coating facilities.

## EXPERIMENTAL

### Sample Preparation

The substrate samples used here were 30 mm × 30 mm niobium coupons produced by electro-discharge machining (EDM) cutting 3 mm thick, RRR>300 sheet material of the type used for cavity fabrication. These samples received 100-150 µm bulk material removal using buffer chemical polishing (BCP) or electropolishing (EP) to remove the damaged layers from the surface. Each sample was treated at 800 °C for 2-3 hours. Samples then received the final removal of 25 μm via EP or BCP. One sample was mechanically polished for the smoothest surface that followed 15 μm EP removal. Nb$_3$Sn thin films were then grown on these samples following a typical coating procedure at Jefferson Lab or Fermilab following typical coating procedures. In this study, we used five samples:

a. MC01 (BCP'ed substrate, coated in FNAL)
b. MC07 (Mechanical polishing (MP) -> EP'ed substrate, coated in JLab)
c. GE70 (EP'ed substrate, coated in FNAL)
d. GE71 (EP'ed substrate, coated in JLab)

___
*work supported by the U.S. Department of Energy, Office of Science, Office of Nuclear Physics under contract DE-AC05-06OR23177 with Jefferson Science Associates, including supplemental funding via the DOE Early Career Award to G. Eremeev. The manuscript has been authored by Fermi Research Alliance, LLC, under Contract No. DE-AC02-07CH11359 with the U.S. Department of Energy, Office of Science, Office of High Energy
† uttar@jlab.org

*Nanoindentation*

Nanoindentation is a widely used technique for measuring the mechanical properties of thin films [14-16]. This technique typically involves indenting the film with a small diamond tip and continuously recording the displacement and load. Nanoindentation equipment allows precise load or displacement control during measurement with small applied forces in nN scales. Fig. 1 illustrates a typical nanoindentation measurement that consists of a three-step process; loading, holding, and unloading. During loading, the load increases with indentation depth consisting of deformation and plastic deformation. In the unloading stage, elastic deformation can be recovered during the unloading that can be used to obtain the film's elastic modulus, hardness, and other mechanical properties.

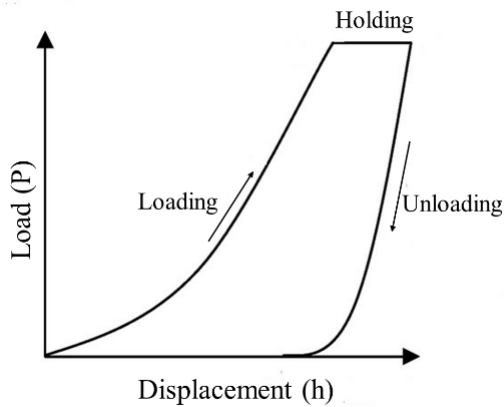

Figure 1: Schematic of the load-displacement curve during a typical nanoindentation measurement.

Nanoindentation measurements were performed on each sample using a Nano Test Vantage instrument (Micro Materials, Wrexham, UK) equipped with a Berkovich diamond indenter at MechAction Lab. The instrument was calibrated before conducting measurements on the Nb$_3$Sn/Nb samples to ensure the lowest noise floor and thermal drift rate. The system and the indenter tip were also validated using fused silica and tungsten reference samples per the ISO 14577 standard. The applied maximum load for each indentation was set to 10 mN for the maximum indent depth below 1/10$^{th}$ of Nb$_3$sn coating thickness to avoid severe substrate effects. A total of 30-50 indentations were performed on each sample, with an indent spacing of 10 μm between adjacent indentations. The loading, holding, and unloading times were set to 5, 2, and 5 s, respectively. The testing parameters and methods followed ASTM E2546 and ISO 14577 standards to ensure the accuracy and reliability of the measurements. Because of the surface roughness of the Nb$_3$Sn surface, see Fig. 2; we only reported 40-60% of the total indentation with consistent results. In the first batch of testing, MC01 and GE70, both coated at the FNAL facility, were tested with ~30 indentations in each sample, out of which ~ 12 were used for the analysis. The other two samples were indented in >50 spots for better statistics, where ~25 indentations were considered for analysis.

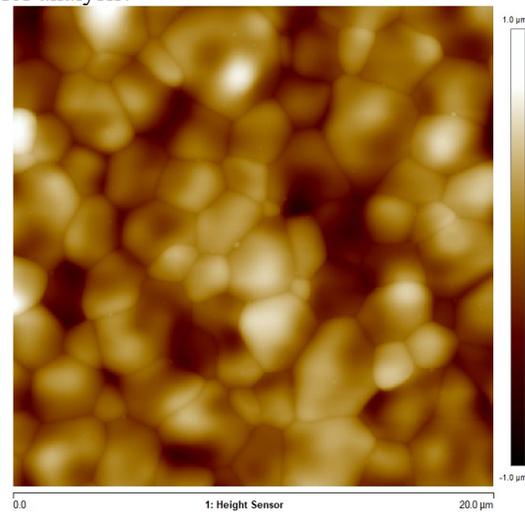

Figure 2: Topography of vapor diffused Nb$_3$Sn from the sample MC07. Note that the roughness is about 1 μm.

During the P-h curve measurement, the indenter is driven into the material producing an impression with a projected area ($A_p$). The indentation hardness, which measures resistance to plastic deformation, can be estimated as $H_{IT} = P_{max}/A_p$, where $P_{max}$ is the maximal load. The Vicker's hardness is defined as $H_v = 94.5 \times H_{IT}$, where $H_{IT}$ and $H_v$ are in GPa and Vickers, respectively. The estimation of Young's modulus, $E_I$, is obtained from the Hertzian theory of contact mechanics [17], which uses the slope of the unload at the maximum displacement point $h_{max}$ (S), $A_p$, modulus and Poisson ratio of the indentor, and Poisson's ratio of the sample. Our calculation is based on the assumption of a Poisson's ratio (ν) of 0.4 for typical Nb$_3$Sn material. Please, see the reference for more details on estimating the modulus value. It should be noted that Young's modulus may vary slightly depending on the assumed Poisson ratio value. Like most metallic materials, yield stress ($\sigma_y$) values are estimated as 1/3$^{rd}$ of the indentation hardness $H_{IT}$.

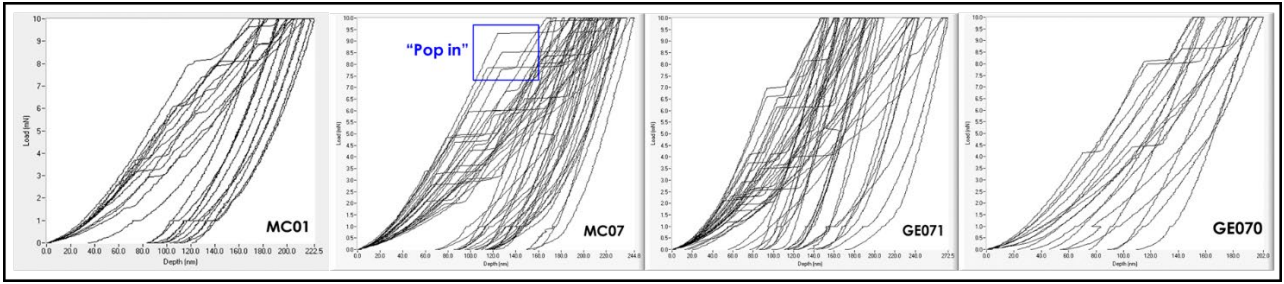

Figure 3: Load-displacement (P-h) curves obtained from nanoindentation of each Nb$_3$Sn-coated sample. Note "pop-in" events in each sample characterized by a distinct drop in the indentation load and an associated discontinuity in the depth of the indenter.

Table 1: Mechanical properties of vapor diffused Nb$_3$Sn on Nb

| Sample | Substrate Preparation | Indentation Hardness $H_{IT}$ (GPa) | Vickers Hardness $H_v$ (Vickers) | Young's Modulus $E_I$ (GPa) | Yield Stress $\sigma_y$ (GPa) | Coating Facility |
|---|---|---|---|---|---|---|
| MC01 | BCP | 10.36 ±1.65 | 979.10±155.80 | 150.06±14.22 | 3.45±0.55 | Fermilab |
| MC07 | MP -> EP | 10.50 ±2.28 | 991.9 ± 215.3 | 164.99±25.71 | 3.50±0.76 | JLab |
| GE070 | EP | 14.40±3.29 | 1360.4 ± 310.9 | 161.2 ± 27.70 | 4.80±1.10 | Fermilab |
| GE071 | EP | 12.82 ±4.55 | 1211.1± 430.0 | 201.92±56.91 | 4.27±1.52 | JLab |
| C-29 (Nb) | BCP | 1.2 ±0.09 | 114.9±8.1 | 116.02±7.35 | 0.41±0.03 | - |

## RESULTS

Ensembles of P-h curves obtained from the indentations of each sample are shown in Fig. 3. Almost all the curves of each sample have shown "pop-in" events on the loading side. Only occasionally, "pop-outs" or "elbows" were observed during the unloading.

The mechanical properties estimated for each sample are tabulated in Table 1. The estimated average among all the samples for $H_{IT}$, $H_v$, $E_I$, and $\sigma y$ are 11.98±1.98, 1135.63±183.83, 169±22.49, and 4.00±0.65 GPa, respectively, where the errors are standard deviation for average estimated values for each sample in Table 1. A SRF cavity grade Nb sample was also characterized; see the P-h indentation curve in Fig.4 using the same measurement instrument to validate the technique. Unlike Nb$_3$Sn, each P-h curve for Nb is more consistent and shows no "pop-in" event, as expected for the soft material. The estimated values for each mechanical parameter are also tabulated in Table 1.

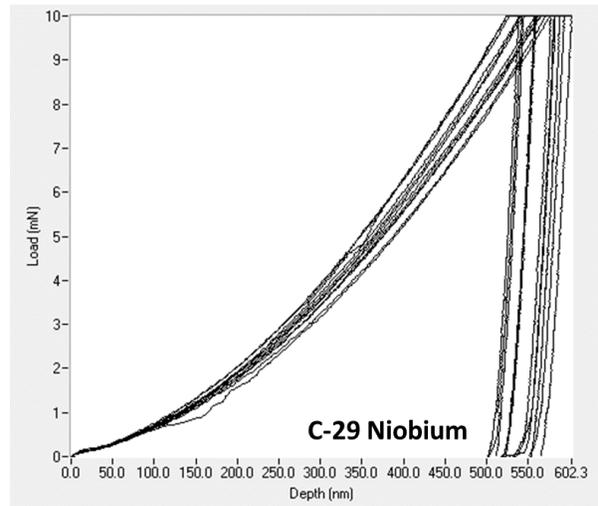

Figure 4: Load-displacement (P-h) curves obtained from nanoindentation of a Nb sample.

## DISCUSSION

The nanoindentation technique differed from the usual tensile tests used to analyze the mechanical characteristics of SRF cavity materials. The measurement was done on a Nb sample to validate the technique. The obtained values for the hardness (1.2±0.09 GPa) and Young's modulus (116.02±7.35 GPa) are within the values typically found in the literature 0.87-1.3 GPa and 105-124 GPa and [18], respectively. The yield strength can be as low as 35-70 MPa for well-annealed Nb to some 100s of MPa for heavily deformed samples [19]. Since the measured Nb sample was not annealed and was not subjected to bulk removal, the indentation was performed on the deformed/ damaged surface layer, likely resulting in a higher value of yield strength.

The observed data show the distinction between soft Nb vs. hard $Nb_3Sn$. The 'pop-in' events observed in $Nb_3Sn$ are most likely because of the generation of micro-cracks during the loading. Similar 'pop-ins' were observed experimentally and linked to the fracture of the brittle film in several studies. Note that we have not observed multiple 'pop-ins' in our experiments but observed single 'pop-ins' as shown in Figure 5. A comparison of the number of 'pop-in' events relative to different loading forces of two samples coated in identical conditions is shown in a histogram in Figure 6, and does not indicate a common correlation between the loading and pop-in in different samples. Note that the measurement values for the hardness from MC01 and MC07 coated in the two different facilities are very similar that is similar to GE070 and GE071. Since each pair of these samples was fabricated from a different batch of materials, more studies are required to see if that has any correlation in resulting in mechanical parameters.

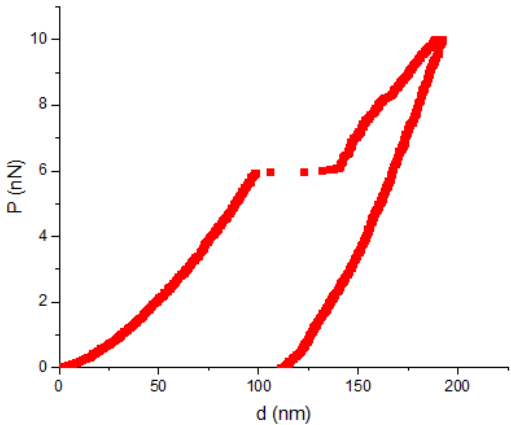

Figure 5: P-d curve for $Nb_3Sn/Nb$ sample with typical single 'pop-in.'

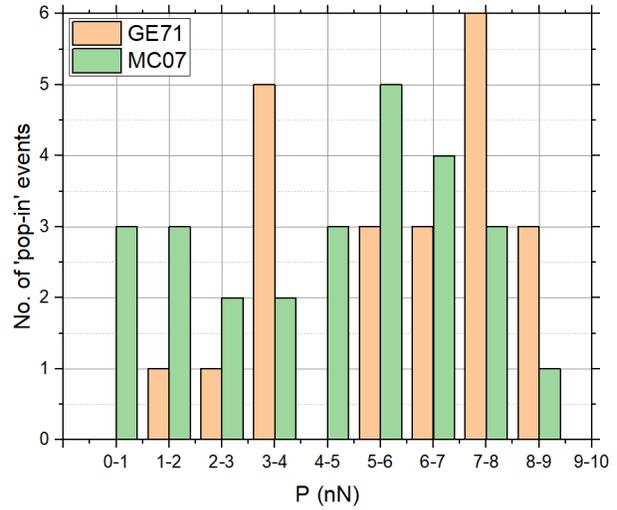

Figure 6: Comparison of 'pop-in' events occurrence at different loadings between sample GE71 and MC07.

## SUMMARY AND OUTLOOK

A set of vapor-diffused $Nb_3Sn$ thin films coated on Nb at JLab and Fermilab coating facilities were examined with the nanoindentation technique, and preliminary data were presented. The 'pop-in' events resulting from the material's cracking show the hard and brittle nature of the material. as these events likely resulted from the initiation and propagation of micro cracks. Despite the surface roughness, we have estimated average mechanical parameters among all the $Nb_3Sn$ samples for $H_{IT}$, Hv, $E_I$, and σy are 11.98±1.98, 1135.63±183.83, 169±22.49, and 4.00±0.65 GPa, respectively. These preliminary values are expected to be valuable in understanding the mechanical limitations for tuning $Nb_3Sn$-coated cavities. These values will be used to simulate the tuning of the $Nb_3Sn$-coated Nb cavity in the near future.

We look forward to using the nanoindentation technique to study the effect of different coating characteristics, such as thickness, grain size, orientation, and grain boundaries while improving the accuracy of the measurement.

## ACKNOWLEDGMENTS

We thank Bo Zhou from MechAction, Inc. for providing measuring our samples and for helpful discussions. Thanks to Eric Lechner and Carrie Baxley for the help with polishing some substrates. We are grateful to Olga Trifimova for her help with the AFM analysis. AFM measurement was done at the Applied Research Center Core Labs, College of Willam & Mary.